\documentclass[11pt,a4paper,reqno]{amsart}
\usepackage{amsmath}
\usepackage[english]{babel}
\usepackage{latexsym}
\usepackage{amssymb}
\usepackage{amscd}
\usepackage{amsgen,amstext,amsbsy,amsopn,amsfonts}
\usepackage{math rsfs}
\usepackage{bm,bbm}
\usepackage{amsthm,epsfig,graphicx,graphics}
\usepackage[latin1]{inputenc}
\usepackage{xspace}
\usepackage{amsxtra}
\usepackage{color}
\usepackage{dsfont}
\usepackage{enumerate}
\usepackage{tikz}
\usepackage{hyperref}
\usetikzlibrary{angles}
\usetikzlibrary{quotes}
\usetikzlibrary{patterns}
\usepackage[font=footnotesize]{caption}
\usepackage[sort&compress,capitalise,nameinlink]{cleveref}

\crefname{section}{\textsection}{\textsection}

\usepackage[deletedmarkup=xout
]{changes}
\definechangesauthor[name={Michele}, color={red}]{michele}
\definechangesauthor[name={Emanuela}, color={blue}]{emanuela}

\usepackage{hyperref}

\usepackage{geometry}
\DeclareMathAlphabet{\mathpzc}{OT1}{pzc}{m}{it}

\makeatletter
\renewcommand*{\@textcolor}[3]{%
  \protect\leavevmode
  \begingroup
    \color#1{#2}#3%
  \endgroup
}
\makeatother

\numberwithin{equation}{section}
\newcommand{\bdm}{\begin{displaymath}}
\newcommand{\edm}{\end{displaymath}}
\newcommand{\bay}{\begin{array}{c}}
\newcommand{\eay}{\end{array}}
\newcommand{\ben}{\begin{enumerate}}
\newcommand{\een}{\end{enumerate}}
\newcommand{\beq}{\begin{equation}}
\newcommand{\eeq}{\end{equation}}
\newcommand{\beqn}{\begin{eqnarray}}
\newcommand{\eeqn}{\end{eqnarray}}

\newcommand{\lf}{\left}
\newcommand{\ri}{\right}

\newcommand{\xv}{\mathbf{x}}

\newcommand{\rv}{\mathbf{r}}

\newcommand{\fv}{\mathbf{F}}
\newcommand{\ev}{\mathbf{e}}
\newcommand{\gamv}{\bm{\gamma}}

\newcommand{\diff}{\mathrm{d}}
\newcommand{\eps}{\varepsilon}

\newcommand{\dist}{\mathrm{dist}}

\newcommand{\as}{\alpha_{\star}}

\newcommand{\nuv}{\bm{\nu}}

\renewcommand{\ss}{\mathsf{s}}

\newcommand{\glf}{\mathcal{E}^{\mathrm{GL}}}
\newcommand{\gle}{E^{\mathrm{GL}}}
\newcommand{\glm}{\psi^{\mathrm{GL}}}

\newcommand{\gldom}{\mathscr{D}^{\mathrm{GL}}}
\newcommand{\aav}{\mathbf{A}}

\newcommand{\aavm}{\mathbf{A}^{\mathrm{GL}}}

\newcommand{\hex}{h_{\mathrm{ex}}}
\newcommand{\theo}{\Theta_0}

\newcommand{\glfk}{\mathcal{G}_{\kappa}^{\mathrm{GL}}}

\newcommand{\glfe}{\mathcal{E}_{\eps}^{\mathrm{GL}}}

\newcommand{\glee}{E_{\eps}^{\mathrm{GL}}}

\newcommand{\doms}{\mathscr{D}_{\star}}

\newcommand{\eones}{E^{\mathrm{1D}}_{\star}}

\newcommand{\fs}{f_{\star}}

\newcommand{\onedom}{\mathscr{D}^{\mathrm{1D}}}

\newcommand{\curl}{\mathrm{curl}}

\newcommand{\anne}{\mathcal{A}_{\eps}}

\newcommand{\disp}{\displaystyle}
\newcommand{\tx}{\textstyle}

\newcommand{\R}{\mathbb{R}}

\newcommand{\E}{\mathcal{E}}

\newcommand{\OO}{\mathcal{O}}

\newcommand{\Om}{\Omega}

\newcommand{\one}{\mathds{1}}

\newcommand{\Hc}{H_{\mathrm{c}1}}
\newcommand{\Hcc}{H_{\mathrm{c}2}}
\newcommand{\Hccc}{H_{\mathrm{c}3}}
\newcommand{\Hstar}{H_{\mathrm{corner}}}

\newcommand{\logi}{|\log \eps| ^{\infty}}

\newtheorem{teo}{Theorem}[section]

\newtheorem{pro}{Proposition}[section]

\newtheorem{asum}{Assumption}
\newtheorem{conj}{Conjecture}

\theoremstyle{remark}
\newtheorem{remark}{Remark}[section]

\newcommand{\fone}{\E^{\mathrm{1D}}}
\newcommand{\fonekal}{\E ^{\rm 1D}_{k,\alpha}}

\newcommand{\eone}{E^{\mathrm{1D}}}

\newcommand{\eoneo}{E ^{\rm 1D}_{0}}

\newcommand{\potkal}{V_{k,\alpha}}

\newcommand{\jv}{\mathbf{j}}

\renewcommand{\leq}{\leqslant}
\renewcommand{\geq}{\geqslant}
\renewcommand{\Im}{\mathrm{Im}}

\newcommand{\corner}{\Gamma_{\beta}(L,\ell)}

\newcommand{\bdo}{\partial \Gamma_{\mathrm{out}}}
\newcommand{\bdi}{\partial \Gamma_{\mathrm{in}}}
\newcommand{\bdbd}{\partial \Gamma_{\mathrm{bd}}}

\newcommand{\osmooth}{\partial \Omega_{\mathrm{smooth}}}

\newcommand{\elle}{\ell_{\eps}}

\newcommand{\ecorr}{E_{\mathrm{corr}}}
\newcommand{\ecorn}{E_{\mathrm{corner},\beta}}
\newcommand{\ecornl}{E_{\mathrm{corner},\beta}(L,\ell)}

\begin{document}

\title{Surface Effects in Superconductors with Corners}

\author[M. Correggi]{Michele Correggi}
\address{Dipartimento di Matematica, Politecnico di Milano, Piazza Leonardo da Vinci, 32, 20133 Milano, Italy.}
\email{michele.correggi@gmail.com}

\date{\today}

\begin{abstract}
We review some recent results on the phenomenon of surface superconductivity in the framework of Ginzburg-Landau theory for extreme type-II materials. In particular, we focus on the response of the superconductor to a strong longitudinal magnetic field in the regime where superconductivity survives only along the boundary of the wire. We derive the energy and density asymptotics for samples with smooth cross section, up to curvature-dependent terms. Furthermore, we discuss the corrections in presence of corners at the boundary of the sample.
\end{abstract}

\maketitle

\tableofcontents

\section{Introduction}\label{sec:intro}

The quantum phenomenon of superconductivity was discovered at the beginning of 1900 by Kamerlingh Onnes and its physics is nowadays quite well understood, at least in the conventional manifestations \cite{Ti}: certain conducting materials may show at low temperature no resistance to the current flow, due to the occurrence of a phase transition and the emergence of a collective behavior of charge carriers. At a microscopic level, the description of superconductivity relies on the Bardeen-Cooper-Schrieffer (BCS) theory \cite{BCS}: the phenomenon is modeled via the formation of weakly-bound pairs of fermions ({\it Cooper pairs}), which in turn combine themselves to give rise to a sort of charge Bose-Einstein condensate. The net effect at the macroscopic level is a sudden drop of resistivity below a certain critical temperature depending on the material. 

\subsection{The Ginzburg-Landau Theory}

The microscopic description of superconductivity provided by BCS theory is not so helpful to extract precise predictions concerning real experiments, given the huge number of degrees of freedom involved. However, close to the critical point for the phase transition, i.e., for temperatures not too far from the critical one, an effective theory of superconductivity is available and it has proven to be very efficient: it is the Ginzburg-Landau (GL) theory \cite{GL}, which was introduced before the proposal of the BCS theory relying only on phenomenological considerations. Only later is was shown that the GL theory emerges in suitable regimes from the BCS microscopic description (see \cite{FHSS} for a rigorous derivation and references therein).

The key idea is that the collective behavior emerging in the material allows to describe the system of charge carriers through a macroscopic wave function $ \psi $, a.k.a. {\it order parameter}, similarly as it can be done for Bose-Einstein condensates. More precisely the physical meaning of such a function is that its square modulus $ |\psi|^2 $ is the relative density of Cooper pairs, i.e., it varies between 0 -- no Cooper pairs and therefore no superconductivity -- and 1 -- all the electrons arranged in Cooper pairs and thus perfect superconductivity. In presence of inhomogeneities, as those induced by an external magnetic fields, $ \psi $ would vary inside the sample itself. 

The free energy of the superconductor is a functional depending on $ \psi $, which is explicitly given by the GL energy 
\begin{equation}
	\label{eq: glfk}
	\glfk[\psi, \textbf{A}]= \displaystyle\int_\Omega \diff\textbf{r} \left\{ |(\nabla +i \aav)\psi|^2-\kappa^2|\psi|^2+ \tx\frac{1}{2}\kappa^2|\psi |^4 \ri\} + \displaystyle\int_{\R^2}\diff\textbf{r}\; |\mbox{curl} \aav - \hex|^2.
\end{equation}
Here we are describing an infinite wire of a conductor with constant and simply connected cross section $ \Omega \subset \R^2 $ and the energy is per unit of longitudinal length. The vector potential $ \aav: \R^2 \to \R^2 $ is associated to the induced magnetic field $ h = \curl \aav $, i.e., the field measured inside and outside the sample. The applied magnetic field is assumed to be directed along the wire and have intensity $ \hex > 0 $. Finally, $ \kappa $ is a parameter depending only on the material and is proportional to the inverse of the London penetration depth. Depending on the value of $ \kappa $, one distinguishes between type-I ($ \kappa < 1 $) and type-II superconductors ($ \kappa > 1 $), which have different kinds of phase transitions. We consider here an extreme type-II superconductor, i.e., $ \kappa \gg 1 $.

Any equilibrium state $ (\psi, \aav) $ is thus a critical point of the GL functional and, in particular, any  ground state, i.e., any minimizer of \eqref{eq: glfk}, is a lowest energy state. In order to discuss the minimization, one has first to specify a suitable minimization domain but it is not so difficult to realize that a natural choice is $ (\psi, \aav) \in H^1(\Omega) \times H^1_{\mathrm{loc}}(\R^2; \R^2) $. There is however a key property of the GL functional, which allows to restrict such a domain (see next \cref{sec: surface} and \cref{sec: corners}): as most of fundamental quantum models, the GL theory is {\it gauge invariant}, i.e., the energy \eqref{eq: glfk} does not change under any gauge transformation $ \psi \to e^{i \phi} \psi $, $ \aav \to \aav - \nabla \phi $, for any $ \phi \in H^1(\R^2) $. Furthermore, all the relevant physical observables are gauge invariant, as, e.g.,  the density $ \lf| \psi \ri|^2 $ or the {\it supercurrent} 
\beq	
	\jv_{\aav}[\psi] : = \tx\frac{i}{2} \lf[ \psi \lf(\nabla - i \aav \ri) \psi^* - \psi^* \lf( \nabla + i \aav \ri) \psi \ri],
\eeq
which is the generalization of the usual current $ \jv[\psi] : = \jv_{\mathbf{0}}[\psi]  = \Im \lf( \psi^* \nabla \psi \ri) $.

\subsection{Response to Applied Magnetic Fields}

The physical phenomenon we plan to study here is the response of the superconducting material to the application of an external magnetic field. It is a quite relevant feature, since, as we are going to see, the presence of intense magnetic fields may affect and eventually destroy superconductivity and therefore, even for practical purposes, any information on whether and how this breakdown occurs is precious. The salient features are known since the works of Abrikosov \cite{Ab} and St. James and De Gennes \cite{SJdG} and it is remarkable that the predictions obtained in the framework of the GL theory are quite accurate:
\begin{itemize}
	\item small magnetic fields do not affect the superconductor and the field can not penetrate the sample ({\it Meissner effect}), i.e., the GL configuration is such that $ |\psi| = 1 $ and $ \aav = 0 $ ({\it perfect superconducting state}), so that $ h = 0 $ inside $ \Omega $;
	\item as soon as a first threshold value $ \Hc $ is crossed, isolated defects ({\it vortices}) appear and $ \psi $ vanishes at the corresponding points, where superconductivity is lost and where the magnetic field in non-zero;
	\item the penetration of the magnetic field occurs at isolated points as long as a second threshold $ \Hcc $ is crossed and the stronger is the field the larger the number of defects, which arrange themselves in regular configurations and eventually in the triangular Abrikosov lattice;
	\item above $ \Hcc $ superconductivity may survive only at short distance from the outer boundary of the sample ({\it surface superconductivity}) and the order parameter $ \psi \simeq 0 $ in the interior of $ \Omega $, where the field has penetrated, i.e., $ h \simeq \hex $ there;
	\item finally, for fields above the third threshold $ \Hccc $, superconductivity is completely destroyed, $ \psi = 0 $ and $ h = \hex $ ({\it normal state}) inside $ \Omega $, so that the material has become again a normal conductor.
\end{itemize}

\subsection{Minimization of the Ginzburg-Landau Energy}

The above behavior has a mathematical counterpart in the GL theory, i.e., the minimization of the GL energy \eqref{eq: glfk}, at least asymptotically when $ \kappa \to + \infty $ and for smooth domains: one can indeed obtain sharp values of the aforementioned critical fields, i.e.,
\beqn
	\Hc & = & C_{\Omega} \log \kappa \lf(1 + o(1) \ri);	\label{eq: Hc}	\\
	\Hcc & = & \kappa^2;	\label{eq: Hcc}	\\
	\Hccc & = & \Theta_0^{-1} \lf(1 + o(1) \ri) \kappa^2,	\label{eq: Hccc k}	
\eeqn
with $ \Theta_0 < 1 $ a universal parameter. The first critical field $ \Hc $ is studied in detail in \cite{SS}, while for the second and third ones we refer to \cite{FH1} and references therein. It is however easy to figure out that, when $ \aav = 0 $ (no applied field), the perfect superconductor state $ \psi = 1 $, $ \aav = 0 $ certainly minimizes the energy, since it minimizes the positive terms of the functional. Similarly, the normal state $ \psi = 0 $, $ h = \hex $ is the lowest energy configuration, when $ \hex $ is very large, e.g., if $ \hex \gg \kappa^2 $, because the last term of the functional \eqref{eq: glfk} is the dominant one.

In this review, the focus is on the surface superconductivity regime, which is identified by applied fiels $ \hex $ between the second and third critical values:
\beq
	\label{eq: range h}
	\Hcc < \hex < \Hccc.
\eeq
In this region, it is more convenient to use different units: we set $ \hex = : b \kappa^2 $ and $ \varepsilon := b^{-\frac{1}{2}} \kappa^{-1} $, so that the GL energy becomes
\begin{equation}
	\label{eq: glf}
	\glfe [\psi, \textbf{A}]= \displaystyle\int_{\Omega} \diff\textbf{r}\; \bigg\{ \bigg| \left ( \nabla + i \frac{\textbf{A}}{\varepsilon ^ 2}\right)\psi \bigg|^2 -\frac{1}{2b\varepsilon ^2}(2|\psi|^2-|\psi|^4)	 \bigg\}+\frac{1}{\varepsilon ^4}\displaystyle\int_{\R^2} \diff\textbf{r}\; |\mbox{curl}\textbf{A} - 1|^2,
\end{equation}
with relevant parameters
\beq
	0 < \eps \ll 1, 		\qquad		b \in \lf(1, b_{3} \ri), \qquad b_{3} = \lim_{\kappa \to + \infty} \kappa^{-2} \Hccc(\kappa),
\eeq
where  the range of values of $ b $ is obviously inherited from \eqref{eq: range h}. Note that, according to \eqref{eq: Hccc k}, one gets $ b_{3} = \theo^{-1} $ for smooth domains (see also below), but a priori the presence of boundary singularities may affect the threshold. 

We denote by $ \glee $ and $ \glm, \aavm $ the ground state energy of \eqref{eq: glf} and any corresponding minimizing configuration (see, e.g., \cite{FH1,SS} for a proof of its existence), i.e.,
\beq
	\label{eq: glee}
	\glee : = \inf_{(\psi, \aav) \in \gldom} \glf[\psi, \aav] = \glf[\glm, \aavm],
\eeq
where $ \gldom : = \lf\{ (\psi, \aav) \in H^1(\Omega) \times H^1_{\mathrm{loc}}(\R^2;\R^2) \: | \: \curl \aav - 1 \in L^2(\R^2) \ri\} $. In the following we also consider a slight variation of the functional \eqref{eq: glf} with a different (smaller) integration domain $ \Omega' \subset \Omega $ and we denote it and its ground state energy as $ \glf[\psi, \aav; \Omega'] $ and $ \glee(\Omega') $, respectively. Any minimizing configuration (actually, any critical point) of $ \glfe $ is such that $ \curl \aavm = 1 $ outside $ \Omega $ and $ \lf( \glm, \aavm \ri) \in \gldom $ solves the GL equations
\beq
	\label{eq: GL eqs}
	\begin{cases}
		- \lf( \nabla + i \tx\frac{\aav}{\eps^2} \ri)^2 \psi = \tx\frac{1}{\eps^2} \lf(1 - |\psi|^2 \ri) \psi,		& \mbox{in } \Omega,		\\
		- \tx\frac{1}{\eps^2} \nabla^{\perp} \curl \aav = \jv_{\aav/\eps^2}[\psi] \one_{\Omega},							& \mbox{in } \R^2,	\\
		\nuv \cdot \lf( \nabla + i  \tx\frac{\aav}{\eps^2} \ri) \psi = 0,									& \mbox{on } \partial \Omega.
	\end{cases}
\eeq

\section{Surface Superconductivity in Smooth Domains}
\label{sec: surface}

Before entering the discussion of the effects on surface superconductivity due to the presence of boundary singularities, we describe the phenomenon in samples with smooth boundary. We thus assume in this Sect. alone that $ \partial \Omega $ is smooth. 

As anticipated, the physics of the surface regime between $ \Hcc $ and $ \Hccc $ was known since the `60s, although those heuristics have been provided with some rigorous basis only recently (see below). The key features are the following:
\begin{itemize}
	\item superconductivity is present only up to distance of order $ \eps $ from the outer boundary, while everywhere else the material is in the normal state;
	\item the magnetic field $ h $ inside the sample is approximately equal to the external one, i.e., there has been almost complete penetration;
	\item in order to compensate the huge magnetic field inside, a strong stationary current develops along the boundary.
\end{itemize}

The mathematical investigation of the surface regime has started in \cite{Pa} and later carried on in \cite{AH,FHP} and the series of works \cite{CR1,CR2,CR3,CDR}, which now provide a detailed rigorous counterpart of the above heuristics for two-dimensional superconductors. Mathematically, the break-down of superconductivity far from the boundary takes the form of the so-called {\it Agmon estimates} (see, e.g., \cite[Chpt. 12]{FH1}), which are suitable nonlinear adaptation of spectral bounds for Schr\"{o}dinger-like operators, i.e.,
\beq
	\label{eq: agmon}
		\displaystyle\int_\Om \diff\textbf{r}\;\exp\lf\{\tx\frac{c(b) \: \mathrm{dist} (\textbf{r},\partial\Om)}{\eps}\ri\} \lf\{|\psi|^2+\eps^2 \lf|\lf(\nabla+i \tx\frac{\textbf{A}}{\eps^2}\ri)\psi\ri|^2\ri\} \; \leq \displaystyle\int_{\{\mathrm{dist}(\textbf{r},\partial\Om)\leq \eps\}} \diff \textbf{r}\; |\psi|^2,
\eeq
where the constant $c(b) $ depends only on the parameter $ b $ and $ c(b) > 0 $ only for $ b > 1 $. Since one can prove a maximum principle for $ \psi $ and show that \cite[Prop. 10.3.1]{FH1}
\beq
	\lf\| \psi \ri\|_{L^{\infty}(\Omega)} \leq 1,
\eeq
the r.h.s. of \eqref{eq: agmon} is $ \OO(\eps) $ and therefore both $ \psi $ and its magnetic gradient must exponentially decay inside $ \Omega $ to compensate for the exploding exponential on the l.h.s.. Furthermore, the vanishing of the constant $ c(b) $ as $ b \to 1^+ $ is the reason why the second critical field is conventionally taken to be equal to $ b = 1 $, since, for smaller values of $ b $, it is possible to show that the decay inside $ \Omega $ is power-law instead of exponential as in \eqref{eq: agmon}. 

We have pointed out that the induced magnetic field $ h = \curl \aav $ is expected to be close to the external one, i.e., $ 1 $ in our units, but this leaves a certain freedom in the choice of $ \aav $. However, in the Coulomb gauge, i.e., requiring that $ \nabla \cdot \aav = 0 $, the vector potential $ \aav $ is defined up to an additive constant, which can be chosen, as we do, in such a way that \cite[Lemma 10.3.2]{FH1}
\beq
	\label{eq: vector potential p}
	\lf\| \aavm - \mathbf{F} \ri\|_{L^{p}(\Omega)} \leq C \eps \lf\| \psi \ri\|_{L^2(\Om)}\|\psi\|_{L^4(\Om)},
\eeq
for any $ p \in [2,+\infty) $ and where
\beq
	\label{eq: fv}
	\fv(\rv) : = \tx\frac{1}{2} \lf( - y, x \ri) = : \tx\frac{1}{2} \rv^{\perp}
\eeq
is the reference vector potential, generating a unit magnetic field.

\subsection{Heuristics}

Let us now establish some heuristics about the order parameter and magnetic field behavior in the regime of surface superconductivity. Thanks to the exponential decay \eqref{eq: agmon} and the estimate \eqref{eq: vector potential p}, one can restrict the minimization of the energy to the boundary layer
\beq
	\label{eq: anne}
	\anne : = \lf\{ \rv \in \Omega \: \big| \: \dist\lf(\rv, \partial \Omega\ri) \leq \eps \elle \ri\},
\eeq
where $ \elle $ can be chosen equal to $ c_0 |\log\eps| $, for a possibly large constant $ c_0 > 0 $, in order to ensure that $ \glm $ is smaller than an arbitrarily large power of $ \eps $ in $ \anne^c $, and fix the magnetic field to be equal to $ \fv $, i.e., $ \gle \simeq \glfe[\glm, \fv; \anne] $. Inside $ \anne $, one can pass to rescaled tubular coordinates via the diffeomorphism 
\beq
	\label{eq: tc}
	\rv(s, t) = : \gamv^{\prime}(s) + \eps t \nuv(s), \qquad		(s,t) \in \lf[0, |\partial \Omega|/\eps\ri] \times \lf[0, \elle\ri], 
\eeq
where $ \rv \in \anne $, $ \gamv: [0, |\partial \Omega|/\eps) \to \partial \Omega $ is a parametrization of the boundary, $ \nuv(s) $ is the inward normal to $ \partial \Omega $ at $ \gamv(s) $ and $ t = \dist\lf(\rv, \partial \Omega\ri)/\eps $.
Exploiting the gauge invariance, the magnetic vector potential can be further locally replaced by
\beq
	\label{eq: magnetic replaced}
	\lf( - \eps t + \tx\frac{1}{2} \eps^2 k(s) t^2 + o(\eps^2) \ri) \ev_{s}
\eeq
i.e., with a purely tangential function, where $ k(s) $ stands for the (rescaled) boundary curvature, defined via
\beq
	\label{eq: curvature}
	\gamv^{\prime\prime}(s) =: k(s) \nuv(s).
\eeq

Neglecting for the moment the corrections induced by the curvature and assuming that there is a separation of variables in the minimization of the effective two-dimensional energy, so that 
\begin{equation}
	\label{eq: op heuristics}
	\glm(\rv) \simeq f(t)  e^{i \alpha s} e^{i \phi_{\eps}(\rv)},
\eeq
where $ f $ is {\it real} and positive, $ \alpha \in \R $ and $ \phi_{\eps} $ is the gauge phase needed to implement the replacement \eqref{eq: magnetic replaced}, then the GL energy becomes 
\beq
	\label{eq: 1d model problem}
	 \glee \simeq \frac{\lf| \partial \Omega \ri|}{\eps} \int^{\elle}_0 \mbox{dt} \left\{ |\partial_t f|^2 + (t+\alpha)^2 f^2 -\frac{1}{2b} (2f^2-f^4)\right\} =: \frac{|\partial \Omega| \fone[f]}{\eps}.
\eeq
We have thus deduced that, if the ansatz is correct, then the modulus of the order parameter must minimize a one-dimensional functional (the energy on the r.h.s. of \eqref{eq: 1d model problem}) depending on the real parameter $ \alpha $, appearing in the phase of $ \glm $. More precisely, we would find that in the boundary layer
\beq
	\label{eq: modulus and current}
	\lf| \glm(\rv) \ri| \simeq f\lf( t \ri),	\qquad		\jv[\glm] \simeq f^2(t) \lf( \frac{|\Omega|}{\eps^2} - \frac{\alpha}{\eps} \ri) \ev_s,
\eeq
where the leading term in the expression of the current is entirely due to the gauge phase in \eqref{eq: op heuristics}. Note that the stationary current in $ \anne $ would then be extremely large (of order $ \eps^{-2} $) and proportional to the volume of $ \Omega $, since we are going to see that $ \alpha = \OO(1) $. Indeed, at this stage, $ \alpha $ is just a parameter appearing in the one-dimensional effective energy, but we should expect that the ground state energy is recovered by optimizing over it, i.e., by finding the optimal phase of $ \glm $.

\subsection{Mathematical Results}

	The above heuristics is correct to leading order but neglects the first order corrections due to the curvature of the boundary, which are already apparent in \eqref{eq: magnetic replaced}, since the second term is $ \OO(\eps|\log\eps|) $ compared to the first one in $ \anne $. A similar correction is generated by the jacobian of the diffeomorphism \eqref{eq: tc}, so that the effective one-dimensional energy locally\footnote{In a disc sample the curvature is constant. Similarly, in a small portion of the boundary layer, we can replace the curvature with its mean value, up to higher order corrections.}  becomes
\begin{equation}
	\label{eq: fonekal}
	\fonekal[f] := \displaystyle\int^{\elle}_0 \mbox{dt} (1 - \eps k t) \left\{ |\partial_t f|^2 + \potkal(t) f^2 -\frac{1}{2b} (2f^2-f^4)\right\},
\end{equation}
where $ k \in \R $ is a parameter playing the role of the constant (or the mean-value of) curvature and
\beq
	\label{eq: potkal}
	\potkal(t) = \frac{1}{(1 - \eps k t)^2} \lf(t + \alpha - \tx\frac{1}{2} \eps k t^2 \ri)^2.
\eeq	
Note that, unlike \eqref{eq: 1d model problem}, the above energy \eqref{eq: fonekal} still depends on $ \eps $ in both the integration domain and the energy density itself. However, setting
\beqn
	\label{eq: eones}
	\eones & : = & \inf_{\alpha \in \R} \inf_{f \in \onedom}  \fone[f];		\\
	\label{eq: eone}
	\eone_{k} & : = & \inf_{\alpha \in \R} \inf_{f \in H^1(\R^+)} \fonekal[f],
\eeqn
where $ \onedom $ is the natural minimization domain for $ \fone $, i.e., $ \onedom = H^1(\R^+) \cap \lf\{ f | \: t f \in L^2(\R^+) \ri\} $, then it is not difficult to see\footnote{We denote by $ \logi $ a large but finite (independent of $ \eps $) power of $ |\log\eps| $, which we do not track down, since it is always multiplied by positive powers of $ \eps $.} \cite[Lemma 2.1]{CR3} that, for any finite $ k \in \R $,
\beq
	\label{eq: eonek expansion}
	\eone_k = \eones + \eps k \ecorr + \OO(\eps^{3/2}\logi),
\eeq
where
\beq
	\label{eq: ecorr}
	\ecorr : = \tx\frac{1}{3} \fs^2(0) \as - \eones,
\eeq
and $ \as, \fs $ is a minimizing pair for the variational problem \eqref{eq: eones}.

We now sum up the main results proven in the aforementioned papers about the ground state of the GL functional in the surface superconductivity regime, i.e., for $ 1 <b < \theo^{-1} $, for 	two-dimensional regular domains:
\begin{itemize}
	\item {\it energy}: combining \cite[Thm. 1]{CR2} with \cite[Lemma 2.1]{CR3} and Gauss-Bonnet theorem, we get
		\beq
			\label{eq: gle asympt smooth}
			\glee = \int_{0}^{\frac{|\partial \Omega|}{\eps}} \diff s \: \eone_{k(s)} + \OO(\eps\logi)
			= \frac{|\partial \Omega| \eones}{\eps} - 2\pi \ecorr + \OO(\sqrt{\eps} \logi),
		\eeq
		where the first expression, although more implicit, allows to get a better remainder term;
		
	\item {\it order parameter}: the heuristics \eqref{eq: modulus and current} is vindicated in \cite[Thm. 2]{CR1}, where it is proven that 
		\beq
			\label{eq: pan smooth}
			\lf\| \lf| \glm(\: \cdot \:) \ri| - \fs(\dist(\: \cdot \:,\partial \Omega)/\eps) \ri\|_{L^{\infty}(\mathcal{A}_{\mathrm{bl}})} = o(1);
		\eeq
		in a suitable annular region $ \mathcal{A}_{\mathrm{bl}} \subset\anne $, containing the outer boundary $ \partial \Omega $;
	\item {\it degree and current}: the above estimate in particular guarantees that in the surface regime $ \glm $ does not vanish on $ \partial \Omega $, since $ \fs(0) > 0 $ for $ 1 < b < \theo^{-1} $; hence the winding number of $ \glm $ along $ \partial \Omega  $ is well defined and can be estimated as \cite[Thm. 3]{CR2}
		\beq
			\label{eq: deg}
			\deg \lf( \glm, \partial \Omega \ri) = \frac{|\Omega|}{\eps^2} - \frac{\as}{\eps}(1 + o(1)),
		\eeq
		in accordance with \eqref{eq: modulus and current};
	\item {\it distribution of superconductivity}: as implied by \eqref{eq: pan smooth}, superconductivity is uniformly distributed all over $ \anne $ to leading order, but the corrections are expected to depend on the curvature: indeed, in \cite[Thm. 1.1]{CR3} it is proven that
	\beq
		\frac{1}{2 b} \int_{D} \diff \rv \: \lf| \glm \ri|^4 = - \eps \eones  \lf| \partial \Omega \cap \partial D \ri|  + \eps^2 \ecorr \int_{\partial \Omega \cap \partial D} \diff \ss \: k(\ss/\eps) + o(\eps^2),
	\eeq
	for any measurable set $ D $ intersecting the boundary with angles $ \pi/2 $; since, for $ 1 < b < \theo^{-1}$, $ \ecorr $ is expected to be positive \cite{CDR}, this means that superconductivity is slightly favored where the curvature is larger.
\end{itemize}

		We now briefly discuss the above results. The two different expansions in \eqref{eq: gle asympt smooth} are a consequence of Gauss-Bonnet theorem, since, by \eqref{eq: eonek expansion},
		\bdm
			\int_{0}^{\frac{|\partial \Omega|}{\eps}} \diff s \: \eone_{k(s)} = \frac{|\partial \Omega|}{\eps} \eones - \eps \ecorr \int_{0}^{\frac{|\partial \Omega|}{\eps}} \diff s \: k(s) + o(1) =  \frac{|\partial \Omega|}{\eps} \eones - 2\pi \ecorr + o(1).
		\edm
		Indeed, for any {\it smooth} planar domain, the integral of the mean curvature along the boundary (recall that $ k(s) $ is the rescaled curvature) equals $ 2 \pi $ times the Euler characteristic, which in turn is equal to $ 1 $. The expansion \eqref{eq: gle asympt smooth} has thus a natural interpretation: each portion of the boundary of a smooth domain contributes to the energy to leading order by a quantity $ \eones/\eps $ per unit lenght, which is independent of the curvature. On the opposite, the first order corrections are determined by the curvature and are proportional to the curvature itself multiplied by $ - \ecorr $.

		We remark that the condition $ 1 < b < \theo^{-1} $ is optimal w.r.t the above features in smooth samples, i.e., the transition to the surface regime occurs at the threshold $ b = 1 $ \cite{FK,FK2} (recall also \eqref{eq: agmon}), while the material goes back to the normal state for applied field larger than $  \theo^{-1} $ \cite{FH3}. Therefore, in proper units the second and third critical fields are given by $ b_2 = 1 $ and $ b_3 = \theo^{-1} $, respectively.
		
		The combination of \eqref{eq: pan smooth} and \eqref{eq: deg} yields a strong evidence that \eqref{eq: op heuristics} is correct, at least to leading order. There is however a freedom in the choice of the phase of $ \glm $ which is not totally resolved: on the one hand, $ \glm $ is always defined up to the multiplication by a constant phase factor, but also the presence of an additional non-trivial phase $ \phi(t) $ depending only on the normal coordinate is not explicitly ruled out by the above results. A simple energetic argument on the other hand shows that this can not be the case: adding a $t-$dependent phase to \eqref{eq: op heuristics} would only increase the energy, since all the terms of the two-dimensional effective functional after the replacement \eqref{eq: magnetic replaced} but the kinetic one are independent of such a phase; moreover the kinetic energy would acquire an additional positive term proportional to $ |\psi|^2 |\phi'(t)|^2 $. 
		
		Hence, the order parameter is characterized by a phase with very rapid oscillations along the boundary (in the $s-$direction), which is constant in the normal direction, and by a modulus which varies only along $ t $. Furthermore, $ \glm$ does not vanish close to the boundary and therefore superconductivity is defect free in that region, which is the most relevant physical consequence of \eqref{eq: pan}. While the structure of $ \glm $ is quite clear inside $ \anne $ or close enough to the boundary, what happens at distances much larger than $ \eps $ from $ \partial \Omega $ is unknown: being $ \glm $ exponentially small in $ \eps $ there, it is very hard to extract information about its zeros and phase, although one expects the presence of a large number of vortices more or less uniformly distributed, so that their winding numbers sum up to reconstruct the tangential phase of $ \glm $ in the boundary layer. A similar behavior occurs for rotating Bose-Einstein condensates, although for quite different physical reasons (see, e.g., \cite{CPRY1,CPRY2}): above a critical value of the angular velocity \cite{CRDY1,CRDY2}, vortices fill the bulk of the condensate \cite{CRv,CY} and arrange in regular structures there, until another critical value is reached. At this threshold vortices are expelled from the bulk \cite{CPRY,CRY,CD}, which is confined to a thin annulus at the boundary, and a giant vortex state is formed, although the behavior in the inner region remains to large extent unknown.
		
		The three-dimensional counterpart of the GL minimization described above has also been studied and some results suggest that an analogous surface regime occurs \cite{FKP,FMP}, although in that case the angle between the applied field and surface of the sample plays an important role. Indeed, for magnetic fields parallel to $ \partial \Omega $, the one-dimensional effective energy $ \eones $ still provides a good approximation of the leading term in the GL energy asymptotics \cite[Thm. 1.5]{FMP}.

\section{Effects of Corners}
\label{sec: corners}

The results presented in \cref{sec: surface} give a precise characterization of the key features of surface superconductivity in domains with smooth boundaries, but in practice it is very hard to distinguish a smooth domain from a less regular one, since defects are always present and, even if a domain may look regular on some scale, on a finer one it may contain corners and other kind of singularities. A typical example of a sample used in experiments is given in \cite[Fig. 1]{NSG} and, at first sight, corners are in fact present at the boundary. Hence, for the consistency of the model, it is important to verify that the predictions concerning superconductivity in the surface regime are robust w.r.t the presence of defects at the boundary. 

Another reason why domains with corners play an important role is the strong similarity one observes with the case of  singular applied magnetic fields and, specifically, {\it magnetic steps} (see \cite{Ass,AK} and in particular \cite[Rmk. 1.9]{AKP}). The jump singularity of the applied field along a curve may indeed produce an effective problem with corners generated by the intersection of the curve with the boundary itself. 

In the rest of the paper, we thus take into account a two-dimensional domain (typically obtained as the cross section of an infinite wire), containing at least one corner, as depicted, e.g., in \cref{fig: domain}. The number of boundary singularities is however assumed to be finite and everywhere else the boundary is smooth. Mathematically, we thus assume that $ \partial \Omega $ is Lipschitz (see \cite[Def. 1.4.5.1]{Gr} for a precise definition) and, more precisely, it is a $ C^{\infty} $ {\it curvilinear polygon}.

	\begin{figure}[!ht]
		\begin{center}
		\begin{tikzpicture}[scale=0.7]
			\draw (0.5,3) to (2,5.5);
			\draw (2,5.5) to[bend right= 4] (4.2,5);
			\draw (6,5.7) to[bend right = 3] (4.2,5);
			\draw (6,5.7) to[bend right= 10] (7,3);
			\draw (7,3) to[bend right= 11] (6.3,0.5);
			\draw (4.5,1.3) to[bend right = 8] (2,0.5);
			\draw (4.5, 1.3) to[bend left = 7] (6.3,0.5);
			\draw (2,0.5) to[bend right= 14] (0.5,3);
		\end{tikzpicture}
		\caption{A domain $ \Omega $ with Lipschitz boundary and finitely many corners on $ \partial \Omega $.}\label{fig: domain}
		\end{center}{}
	\end{figure}
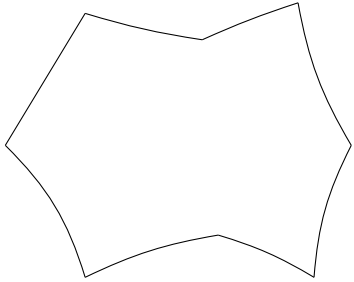

	\begin{asum}[Boundary with corners]
		\label{asum: boundary}
		\mbox{}	\\
		We assume that the boundary $ \partial \Omega $ is a smooth curvilinear polygon and its set  $ \Sigma : = \lf\{ \rv_1, \ldots, \rv_N \ri\} $ of corners is non empty but finite and given by $ N $ points. We denote by $ \beta_j$ the angle of the $j-$th corner (measured towards the interior) and by $ s_j $ its boundary coordinate.
	\end{asum}

	Hence, the unit vector normal to the boundary is defined everywhere but at the corner points, where it jumps from one direction to another. Similarly, the curvature is bounded everywhere but it has jump singularities in the correspondence of corners. 
	
	The minimization problem to consider is then the same as in \eqref{eq: glee} where the GL functional is given by \eqref{eq: glf}. In spite of some minor technical differences and natural adaptations (see, e.g., \cite[Appendix B]{CG2}), all the features of the minimization in smooth domains hold true also in presence of corners. We thus study in the following the asymptotic behavior of $ \glee $ and $ \glm $, as $ \eps \to 0  $, for $ b > 1 $. It is indeed not so difficult to see that the Agmon bounds \eqref{eq: agmon}  applies also to samples with corners (see, e.g., \cite{BNF}) and therefore the second critical field is certainly smaller or equal to $ 1 $. Furthermore, being the nucleation of vortices a bulk phenomenon, one expects that the regime around the first critical field is unaffected by boundary singularities. Hence, the first non-trivial effects of corners should appear for 
	\bdm
		b > 1
	\edm
	and that is the regime we are going to explore.
	
	\subsection{State of the Art}
	\label{sec: art}
	
	Since the second critical field is expected to coincide with the one for smooth samples, one may think that the third critical field too is not influenced by the presence of corners, but actually here a major difference takes place. As shown in various works \cite{BNF,HK,Ja,JRS,Pa2}, in presence of one corner of acute angle $ \beta $, there is a shift of the third critical field (see in particular \cite[Thm. 1.4]{BNF}), whose value becomes asymptocally
\beq
	\label{eq: Hccc}
	\Hccc = \frac{1}{\mu(\beta) \eps^2}
\eeq	
yielding $ b = \mu(\beta)^{-1} $, where
\beq
	\mu(\beta) : = \inf \mathrm{spec}_{L^2(W_\beta)} \lf( - \lf(\nabla + \tx\frac{1}{2} i \xv^{\perp} \ri)^2 \ri),
\eeq
$ W_{\beta} $ being an infinite sector of angle $ \beta $. The above results however are subjected to the assumption (see \cite[Chpt. 8.2]{Ra})
\beq
	\label{eq: conditional}
	\mu(\beta) < \theo,
\eeq
namely, the fact that the ground state of a magnetic Schr\"{o}dinger operator with uniform magnetic field in an infinite sector $ W_{\beta} $ of opening angle $ 0 < \beta < \pi  $ is {\it strictly smaller} than in the case of a half-plane (where the ground state energy equals $ \theo $). The above inequality however is proven only for angles $ 0 < \beta < \pi/2 + \epsilon $ \cite{Bo,Ja,ELP}, although numerical simulations \cite{ABN, BNDMV, Bo,BD,ELP} suggest that it should hold true for any acute angle (see also \cite[Rmk. 1.1]{BNF}). In fact, $ \mu(\beta) $ is expected to be monotone in $ \beta $ between $ 0 $ and $ \pi $ and remain constantly equal to $ \theo $ for $ \pi \leq \beta \leq 2\pi $, so that only acute angles would play a role and, among those, the smallest one determines the threshold for the transition to the normal state.

Furthermore, assuming \eqref{eq: conditional}, the loss of superconductivity occurs in a different way compared to smooth domains: instead of observing a direct transition between a surface state and the normal one, there is an intermediate regime where superconductivity survives near the corners (actually, near the one with smallest opening angle) and the order parameter is exponentially small everywhere else. More precisely, it can be shown \cite[Thm. 1.6]{BNF} that, for $ \theo^{-1} < b < \mu(\beta)^{-1} $, the Agmon estimate \eqref{eq: agmon} holds true with the distance from the set of corners in place of the distance from the boundary. In fact, this last result has been recently improved in \cite{HK}: in presence of several corners so that \eqref{eq: conditional} holds, one can associate each one with a corresponding critical value $ \mu(\vartheta_j)^{-1} $, marking the transition from concentration close to one corner to concentration to another one, until the corner with smallest angle is reached. 

This behavior near the transition to the normal state suggests that, in presence of corners, the surface regime does not cover the whole parameter interval $ (b_2, b_3) $, but rather a smaller region $ (b_2, b_{\mathrm{corner}}) $ with $ b_{\mathrm{corner}} < b_3 $. In other words, we should expect the appearance of a {\it new critical field} $ b_{\mathrm{corner}} $, marking the transition from surface to corner superconductivity.

A key contribution towards the investigation of such a new phase transition was given in \cite{CG}, where the effects in surface superconductivity due to the presence of corners are studied to leading order. It is indeed shown that superconductivity is basically not affected by corners in that approximation and, moreover, the parameter region where one observes the surface behavior contains the parameter interval $ (1, \theo^{-1}) $, so leading to the strong conjecture that $ b_2 = 1 $ and the transition from boundary to corner concentration occurs at $ b_{\mathrm{corner}} = \theo^{-1} $, which in the original units would become
\beq
	\Hstar = \frac{1}{\theo \eps^2}.
\eeq

The precise result proven in \cite[Thm 1.1]{CG} is that, in a domain satisfying \cref{asum: boundary}, if $ 1 < b < \theo^{-1} $, then
\beq
	\label{eq: glee asympt CG}
	\glee = \frac{|\partial \Omega| \eones}{\eps} + \OO(|\log\eps|^2),
\eeq
i.e., the leading term in the GL energy asymptotics coincides with the one in \eqref{eq: gle asympt smooth}, although the remainder is much worse. More importantly, the order parameter is approximately uniformly distributed along the boundary (again to leading order) and
\beq
	\lf\| \lf|\glm( \: \cdot \:) \ri|^2 - \fs^2\lf( \dist( \: \cdot \:, \partial \Omega)/\eps \ri) \ri\|_{L^2(\Omega)} = \OO(\eps|\log\eps|),
\eeq
so substantiating the conjecture about the value of the new critical field $ \Hstar $.

\subsection{Effective Model}
\label{sec: effective}

Before stating our main results, we have to introduce the model problem providing the energy correction due to corners. Heuristically, it is obtained by blowing-up the corner region, so that it becomes an infinite domain with straight boundaries (see \cref{fig: corner}) and subtracting from the GL energy with fixed magnetic field equal to $ \fv $ (recall \eqref{eq: fv}) the energy contribution of the straight portion of the boundary, i.e., $ \eones $ times its length.

Let us first describe in more details the corner region $ \corner $ as in \cref{fig: corner}: the opening angle with vertex $ V $ is $ \beta $ and the side lengths
\beq
	 L =  |\overline{AV}| = |\overline{BV}|,	\qquad	  \ell = |\overline{AC}| = |\overline{EB}|.
\eeq
We always assume that
	\beq
		\label{eq: L ell condition}
		\ell \leq \tan\lf( \tx\frac{\beta}{2}\ri) \, L,
	\eeq
	so that the region has the shape of \cref{fig: corner} and we have
	\beq
		\label{eq: corner}
		\corner : = \lf\{ \rv \in \R^2 \: \big| \: \dist\lf(\rv, \bdo \ri) \leq  \ell \ri\}.
	\eeq
	Here we have denoted by $ \bdo $ the outer boundary\footnote{We often use polar coordinates $ (\varrho, \vartheta) \in \R^+ \times [0, 2\pi) $, with center in the vertex $ V $.}
	\beq
		\label{eq: bdo}
		\bdo : = \lf\{ \rv \in \R^2 \: \big| \: \rv = (\varrho, 0), 0 \leq \varrho \leq L \ri\} \cup  \lf\{ \rv \in \R^2 \: \big| \: \rv = (\varrho, \beta), 0 \leq \varrho \leq L \ri\},
	\eeq
	while the other two components of $ \partial \corner $ are denoted by $ \bdi $ and $ \bdbd $, where $ \bdi : = \lf\{ \rv \in \partial \corner \: \big| \: \dist\lf(\rv, \bdo \ri) = \ell \ri\} $.

	\begin{figure}[!ht]
		\begin{center}
		\begin{tikzpicture}[scale=0.75]
			\draw (0,0) -- (1,2);
			\draw (1,2) -- (1.5,3);
			\draw (1.5,3) -- (2,4) -- (2.5,3);
			\draw (2.5,3) -- (3,2);
			\draw (3,2) -- (4,0);
			\draw (0,0) -- (1,-0.5);
			\draw (4,0) -- (3,-0.5);
			\draw (1,-0.5) -- (2,1.7);
			\draw (3,-0.5) -- (2,1.7);
			\node at (2,4.5) {{\footnotesize $V$}};
			\node at (-0.5, 0) {{\footnotesize $A$}};
			\node at (4.5, 0) {{\footnotesize $B$}};
			\node at (1.5, -0.5) {{\footnotesize $C$}};
			\node at (2.5, -0.5) {{\footnotesize $E$}};
			\node at (2,1) {{\footnotesize $D$}};
		\end{tikzpicture}
		\hspace{0,5cm}
		\begin{tikzpicture}[scale=0.45]
			\draw[->] (0.5,2) to (13.5,2);
			\draw[->](10.5,0.5) to (10.5,9);
			\draw (2.4,2) to (2.4,3.6);
			\draw (2.4,3.6) to (6.9,3.6);
			\draw (4,7) to (6.9,3.6);
			\draw (4,7) to (5.5, 8);
			\draw (5.5,8) to (10.5,2);
			\node at (11.4,1.5) {{\footnotesize $x$}};
			\node at (9.9,8.7) {{\footnotesize $y$}};
		\end{tikzpicture}
		\caption{The corner region $ \corner $ and the associated coordinate system. The opening angle $ \widehat{AVB} $ is equal to $ \beta $ and the side lengths are $ |\overline{AV}| = |\overline{VB}| = L  $ and $  |\overline{AC}| = |\overline{EB}| = \ell $.}\label{fig: corner}
		\end{center}{}
	\end{figure}
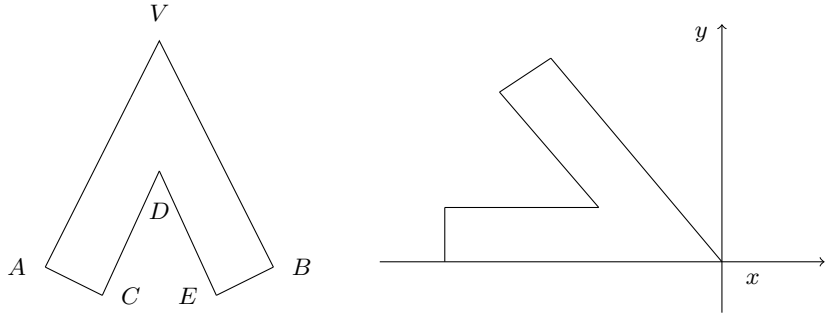

The corner effective energy is given by
	\beq
		\label{eq: ecornl}
		\ecornl : = - 2 L \eoneo(\ell) + \inf_{\psi \in \doms(\corner)} \glf_{1}\lf[\psi, \fv; \corner\ri],
	\eeq
	where we recall that $ \glf_{1}\lf[\psi, \fv; \corner\ri] $ stands for the GL energy \eqref{eq: glf} with $ \eps = 1 $, magnetic field $ \aav = \fv $ (w.r.t. a system of coordinates with the origin in $ V $, as in \cref{fig: corner}) and integration domain given by $\corner $ (a similar functional but on a different domain is introduced in \cite[Eq. (1.11)]{BNF} as the effective energy in an infinite sector). The one-dimensional energy $ \eoneo(\ell) $ is a variant of $ \eones $ defined in \eqref{eq: eones}, differing from it only by the integration domain, which is the finite interval $ [0, \ell] $, i.e., explicitly (for a detailed discussion of the corresponding minimization problem we refer to \cite[Appendix A.3]{CG2})
	\beq
		\label{eq: eoneo}
		\eoneo(\ell) : = \inf_{\alpha \in \R} \inf_{f \in H^1([0,\ell])} \int^{\ell}_0 \mbox{dt} \left\{ |\partial_t f|^2 + (t+\alpha)^2 f^2 -\frac{1}{2b} (2f^2-f^4)\right\},	
	\eeq
	Finally, the minimization domain contains regular functions with suitable boundary conditions along $ \bdbd $ and $ \bdi $, meant in trace sense:
	\beq
		\label{eq: doms}
	 	\doms(\corner) : = \lf\{ \psi \in H^1(\corner) \: \big| \: 
	 	\lf. \psi\ri|_{\bdbd \cup \bdi} = \psi_{\star} \ri\}.
	\eeq
	where the function $ \psi_{\star} $ is expressed in tubular coordinates $ (s,t) \in [-L,L]\times [0,\ell] $, with $ t = \dist(\rv, \bdo) $, as
	\beq
		\label{eq: psi star}
		\psi_{\star}(\rv(s,t)) : = f_0(t) \exp \lf\{ i \alpha_0 s - \tx\frac{1}{2} i st \ri\}.		
	\eeq
	Note that far from the corner and, specifically, in each of the rectangular domains $ R_{\pm} $ with
	\beq
		R_{+} \cup R_- = \lf\{ \rv \in \corner \: \big| \: \dist(\rv, \bdo) + \dist(\rv,\bdi) = \ell \ri\},
	\eeq 
	the tubular coordinates $ (s,t) $ actually define a global diffeomorphism, so that $ R_{\pm} $ are represented as the sets $ [L - L/\tan(\beta/2), L] \times [0,\ell] $ and $ [-L,-L + L/\tan(\beta/2)] \times [0,\ell] $, respectively.
	
As anticipated, the heuristics behind the expression \eqref{eq: ecornl} is that the energy contribution of the corner is obtained by minimizing in a region of size $ \sim \eps $ around it a GL functional with fixed magnetic field $ \fv $ and subtracting the surface energy of the smooth portion of the boundary, which in the domain $ \corner $ equals $ \eones |\bdo| = 2 L \eones $. In this respect, it is important to remark that the boundaries $ \bdi $ and $ \bdbd $ are not expected to provide any surface energy contribution because of the conditions in \eqref{eq: doms}: the modulus of the function $ \psi_{\star} $ indeed depends only on the distance from the outer boundary $ \bdo $, which is thus the only relevant one for surface superconductivity. In the heuristic interpretation of the region $ \corner $, both boundaries $ \bdi $ and $ \bdbd $ are in fact fake ones: the inner region $ \bdi $ does not play any role because the order parameter gets exponentially small there (indeed, $ \psi_{\star} $ is exponentially small in $ \ell $ on $ \bdi $), while the boundary $ \bdbd $ is an artificial one, introduced only to separate the corner region from the smooth part of $ \partial \Omega $.

The reason why the energy $ \ecornl $ can not vanish is that, in order to give the correct surface energy, the minimizer must be close to the function $ \psi_{\star} $ is both rectangular regions $ R_{\pm} $, but the phase $ - i \as s $ shows a jump singularity along the bisectrix of the angle in the vertex. Hence, the phase of the minimizer must be a genuine two-dimensional function connecting the two values of $ - i \as s $ on the two sides of $ \corner $. This in turn implies that $ \psi_{\star} $ itself is a two-dimensional function, i.e., there is no separation of variables, which yields a net contribution to the energy of pure geometric nature above the surface term. In fact, we expect that the phase of any minimizer of \eqref{eq: ecornl} becomes singular along the bisectrix in the limit $ L, \ell \to + \infty $, which is another analogy with the case of magnetic steps \cite{Ass,AKP}.

	Since $ L $ and $ \ell $ are obtained by rescaling of the tangential and normal lengths of the corner region and are therefore of order $ |\log\eps| \gg 1 $ in the original units, the model energy of the corner must be obtained in the limits $ L, \ell \to + \infty $.

	\begin{pro}[Corner energy]
		\label{pro: ecorn}
		\mbox{}	\\
		Let $ \corner $ be the domain depicted in \cref{fig: corner}. Then, for any $ 1 < b < \theo^{-1} $ and $ \beta \in (0, 2\pi) $, there exists finite the limit
		\beq
			\label{eq: ecorn def}
			\lim_{L \to + \infty} \lim_{\ell \to + \infty} \ecornl = : \ecorn.		
		\eeq
	\end{pro}

	As stated in the above \cref{pro: ecorn}, the energy $ \ecorn $ is bounded, but, contrary to what is expected, it might as well be zero, namely we do not have a direct proof that it does not vanish. The question whether this is the case or not is strictly related to the dependence of $ \ecorn $ on the angle $ \beta $ and therefore to its explicit expression. We now discuss this point further: although we do not have a definite answer, we are going to motivate a conjecture about the expression of $ \ecorn $, also consistent with its non-vanishing.

	The starting point of the argument is the explicit form of the energy correction for smooth domains: as we have seen, such an energy can be expressed as
	\beq
		\label{eq: smooth gb}
		-\eps\ecorr \int_{0}^{|\partial \Omega|/\eps} \diff s \: k(s) = -2 \pi \ecorr,
	\eeq
	up to higher order terms. In the above identity we have applied the Gauss-Bonnet theorem, which holds true also for domains with corners along the boundary: the statement however has to be suitably modified, i.e., one has the identity
	\beq
		\label{eq: gb corners}
		\eps \int_{\partial \Omega_{\mathrm{smooth}}} \diff s \: k(s) + \sum_{j =1}^N (\pi - \beta_j) = 2 \pi,
	\eeq
	where we have denoted by $ \partial \Omega_{\mathrm{smooth}} $ the smooth part of the boundary, i.e., the first integral can be explicitly written as
	\bdm
		\eps \int_{\partial \Omega_{\mathrm{smooth}}} \diff s \: k(s) = \eps \sum_{j = 1}^N \int_{s_j}^{s_{j+1}} \diff s \: k(s),
	\edm
	with $ s_j $, $ j = 1, \ldots, N $, the curvilinear coordinates of the corners and the convention that $ s_{N+1} = s_1 + |\partial \Omega|/\eps $. Equivalently, one could integrate $ k(s) $ over the whole boundary, but the integral has then to be meant in Lebesgue sense: recall that the curvature $ k(s) $ is bounded and with finitely many jump singularities, so that it is integrable in Lebesgue sense and its integral coincide with the expression above. 
	
	Formally, the identity \eqref{eq: gb corners} can be reformulated by assuming that the curvature contains in fact delta singularities supported at the corners, i.e., replacing $ k(s) $ with
	\bdm
		k(s) + \eps^{-1} \sum_{j=1}^N (\pi - \beta_j) \delta(s-s_j).
	\edm
	Hence, we may think that the first order correction to the energy has the same form as for smooth domains, where however the above replacement of the curvature has been done. This would give the energy contribution 
	\bdm
		- \eps \ecorr \int_{\partial \Omega_{\mathrm{smooth}}} \diff s \: k(s) - E_{\mathrm{corr}} \sum_{j =1}^N (\pi - \beta_j).
	\edm
	If we take seriously such an argument and compare the energy above with the one defined in \eqref{eq: ecorn def}, we are lead to conjecture that $ \ecorn = - E_{\mathrm{corr}} (\pi - \beta) $, since in $ \corner $ there is only one corner with angle $ \beta  $ and the curvature vanishes everywhere.

	\begin{conj}[Corner energy]
		\label{conj: ecorn}
		\mbox{}	\\
		For any $ 1 < b < \theo^{-1} $ and $ \beta \in (0,2\pi) $, one has
		\beq
			\label{eq: ecorn conj}
			\ecorn = -(\pi - \beta) \ecorr.
		\eeq
	\end{conj}

	\begin{remark}[Linear regime $b \to \theo^{-1} $]
		\mbox{}	\\
		When $ b \to \theo^{-1} $, the linear regime must be recovered, namely the model problem becomes the Schr\"{o}dinger operator with uniform magnetic field in an infinite sector. As already mentioned in \cref{sec: art}, this is expected to introduce a strong asymmetry between corners with acute and obtuse angles \cite[Rmk. 1.1]{BNF}: the numerical simulations \cite{ABN,BNDMV} on the ground state of the Schr\"{o}dinger operator suggest that $ \mu(\beta) $ is a monotone function of $ \beta \in (0, \pi) $ and then remains constantly equal to $ \theo $ for angles larger than $ \pi $. Therefore, corners with obtuse angles would not play any role nor affect the energy asymptotics. On the opposite, in \eqref{eq: ecorn conj}, the correction to the energy is present irrespective of the corner angle. Furthermore, it is known that in both the surface and linear regimes points with large curvature attracts superconductivity, which becomes stronger there. This is correlated with the sign of $ \ecorr $, which is known to get positive, as $ b \to \theo^{-1} $ (for $ 1 < b < \theo^{-1} $, it is verified numerically in \cite{CDR}). This is however not in contradiction with \eqref{eq: ecorn conj}, since the concentration is a first order effect in the surface superconductivity regime and the limit $ b \to \theo^{-1} $ is singular, since both $ \eones $ and $ \ecorr $ vanish, and therefore the scheme does not apply any longer.
	\end{remark}
	
	\begin{remark}[Almost flat angles]
		\mbox{}	\\
		Despite a lack of a proof of \cref{conj: ecorn}, an interesting result which makes it stronger is proven in the companion paper \cite{CG3}: we show that in a domain with a corner of opening angle $ \pi \pm \delta $, with $ 0 < \delta \ll 1 $, the corner energy can be expanded as follows:
		\beq
			\ecorn = \mp \delta \ecorr + \OO(\delta^{4/3}) + \OO(\ell^{-\infty}),
		\eeq
		i.e., the conjecture is confirmed to leading order. More importantly, this result also shows that the corner energy is not zero, at least for angle suitably close to $ \pi  $.	
	\end{remark}

\subsection{Main Results}

We are now able to state our main results concerning the effects of corners to the asymptotics of the GL energy.

	\begin{teo}[GL energy asymptotics]
		\label{teo: gle asympt}
		\mbox{}	\\
		Let $\Om\subset\R^2$ be any bounded simply connected domain satisfying \cref{asum: boundary}. Then, for any fixed $ 1< b < \theo^{-1} $, as $\eps \to 0$, it holds
		\beq
			\label{eq: gle asympt}
			\glee = \disp\frac{|\partial\Om| \eones}{\eps} - \eps \ecorr \int_{0}^{|\partial \Omega|/\eps} \diff s \: k(s) + \sum_{j = 1}^N E_{\mathrm{corner}, \beta_j} + o(1), 
		\eeq
		where the integral of the curvature is meant in Lebesgue sense, or, equivalently, performed over the smooth part of $ \partial \Omega $.
	\end{teo}

	\begin{remark}[Critical fields]
		\label{rem: critical fields}
		\mbox{}	\\
		The parameter region $ 1 < b < \theo^{-1} $ is expected to be sharp for the result above, i.e., it should identify the surface superconductivity regime in presence of corners, exactly as in the case of smooth domains. Hence, it would follow that, on the one hand, $ b_2 = 1 $ and therefore
		\beq
			\Hcc = \frac{1}{\eps^2}.	
		\eeq
		This is also vindicated by the results proven in \cite{FK}. On the other hand, the transition from boundary to corner behavior should occur at $ b_{\mathrm{corner}} = \theo^{-1} $ and thus
		\beq
				\Hstar = \frac{1}{\theo \eps^2}.
		\eeq
		In presence of $ N $ corners such that $ \mu(\vartheta_j) < \theo $, the critical field above would then be followed by $ N $ thresholds \cite[Rmk. 1.4]{HK}, marking the concentration of superconductivity close to the corresponding corner until the normal state is recovered:
		\beq
			\Hstar < H_{\mathrm{corner},1}  \leq \ldots \leq H_{\mathrm{corner},N-1} \leq H_{\mathrm{corner},N} = \Hccc,
		\eeq
		with
		\beq
			 H_{\mathrm{corner},j} = \frac{1}{\mu(\beta_j) \eps^2},		\qquad 	\mbox{for } 1 \leq j \leq N.
		\eeq
	\end{remark}

		\begin{remark}[$ L^2 $ estimate of the order parameter]
		\label{rem: refined l2 estimate}
		\mbox{}	\\
		As usual, the estimate of the energy asymptotics yields some information on the behavior of the order parameter and it is not difficult to show that \eqref{eq: gle asympt} allows to improve the error \eqref{eq: glee asympt CG} as
		\beq
			\label{eq: refined l2 estimate}
			\lf\| \lf| \glm(\: \cdot \:) \ri|^2 - {\fs}^2\lf( \dist(\: \cdot \:, \partial \Omega)/\eps \ri) \ri\|_{L^2(\Omega_{\mathrm{smooth}})} = o(\eps|\log\eps|),
		\eeq
		where $ \Omega_{\mathrm{smooth}} : = \lf\{ \rv \in \Omega | \dist(\rv, \Sigma) \geq c_2 \eps |\log\eps| \ri\} $, for some large constant $ c_2 $ independent of $ \eps $.
	\end{remark}
	
	According to the above \cref{rem: refined l2 estimate}, the order parameter and thus superconductivity is to leading order uniformly distributed in $ L^p $ sense in $ \anne $. This however leaves room for isolated defects where $ \glm $ goes to zero on a suitable short scale. In order to rule out the occurrence of such a points or small regions where superconductivity is lost, one would need to strengthen \eqref{eq: refined l2 estimate} and extract an $L^{\infty} $ bound of $ \glm $ analogous to \eqref{eq: pan smooth}. There is however a main obstruction towards such a proof related to the lack of information on the minimizer of the effective model discussed in the previous \cref{sec: effective}, which allows to investigate the detailed structure of $ \glm $ only far from corners.

	\begin{pro}[GL order parameter]
		\label{pro: pan}
		\mbox{}	\\
		Under the same assumptions of \cref{teo: gle asympt},
		\beq
			\label{eq: pan}
			\lf\| |\glm(\rv)| - {\fs}(0) \ri\|_{L^{\infty}(\osmooth)} = o(1).
		\eeq
		where $ \osmooth : = \lf\{ \rv \in \partial \Omega \: \big| \: \dist(\rv, \Sigma) \geq c_2 \eps |\log\eps| \ri\} $.
	\end{pro}
	
	\begin{remark}[Uniform distribution of superconductivity]
		\label{rem: uniform}
		\mbox{}	\\
		The bound \eqref{eq: pan} is stated only along the boundary, but can in fact be extended to a layer as in \eqref{eq: pan smooth}, so providing a stronger insight on the uniformity of superconductivity along the boundary region. Due to the impossibility of ruling out zeros of  $\glm $ close to the corners, we can not translate \eqref{eq: pan} into an estimate of the current along $ \partial \Omega $.
	\end{remark}

\end{document}